\documentclass[%
 aip,
 amsmath,amssymb,
 reprint,%
]{revtex4-2}

\usepackage{graphicx}
\usepackage{dcolumn}
\usepackage{bm}

\usepackage[utf8]{inputenc}
\usepackage[T1]{fontenc}
\usepackage{mathptmx}

\begin{document}

\title[MD Simulations of Charged Binary Mixtures Reveal a Generic Relation Between High- and Low-Temperature Behavior]{MD Simulations of Charged Binary Mixtures Reveal a Generic Relation Between High- and Low-Temperature Behavior}

\author{L. Hecht}
 \email{hecht@fkp.tu-darmstadt.de}
 \affiliation{ 
 	Institut für Physik kondensierter Materie, Technische Universität Darmstadt, Hochschulstr. 8, 64289 Darmstadt, Germany
 }
\author{R. Horstmann}
\affiliation{ 
	Institut für Physik kondensierter Materie, Technische Universität Darmstadt, Hochschulstr. 6, 64289 Darmstadt, Germany
}
\author{B. Liebchen}
\affiliation{ 
	Institut für Physik kondensierter Materie, Technische Universität Darmstadt, Hochschulstr. 8, 64289 Darmstadt, Germany
}
\author{M. Vogel}
 \affiliation{ 
Institut für Physik kondensierter Materie, Technische Universität Darmstadt, Hochschulstr. 6, 64289 Darmstadt, Germany
}%

\date{11 December 2020}

\begin{abstract}
	Experimental studies of the glassy slowdown in molecular liquids indicate that the high-temperature activation energy $E_{\infty}$ of glass-forming liquids is directly related to their glass transition temperature $T_{\text{g}}$. To further investigate such a possible relation between high- and low-temperature dynamics in glass-forming liquids, we analyze the glassy dynamics of binary mixtures using molecular dynamics (MD) simulations. We consider a binary mixture of charged Lennard-Jones particles and vary the partial charges of the particles, and thus, the high-temperature activation energy and the glass transition temperature of the system. Based on previous results, we introduce a phenomenological model describing relaxation times over the whole temperature regime from high temperatures to temperatures well inside the supercooled regime. By investigating the dynamics of both particle species on molecular and diffusive length scales along isochoric and isobaric pathways, we find a quadratic charge dependence of both $E_{\infty}$ and $T_{\text{g}}$, resulting in an approximately constant ratio of both quantities independent of the underlying observable, the thermodynamic ensemble, and the particle species, and this result is robust against the actual definition of $T_{\text{g}}$. This generic relation between the activation energy and the glass transition temperature indicates that high-temperature dynamics and the glassy slowdown are related phenomena, and the knowledge of $E_{\infty}$ may allow to approximately predict $T_{\text{g}}$.
\end{abstract}

\maketitle

\section{Introduction}
\label{sec:Introduction}

Glass-forming liquids can be in the liquid phase even below the melting point $T_{\text{m}}$. In this supercooled regime, a dramatic slowdown of dynamics occurs if the temperature $T$ is decreased, resulting in a sharply growing structural relaxation time $\tau$ and viscosity $\eta$. At the glass transition temperature $T_{\text{g}}$, the structural relaxation time becomes larger than all relevant experimental time scales and the liquid falls out of equilibrium. In experiments, the glass transition temperature is often defined as \text{$\tau\left(T_{\text{g}}\right)=1000\,\text{s}$}.\cite{Cavagna_PhysRep_2009} Interestingly, experimental studies propose the existence of a simple and generic relation between high-temperature dynamics and $T_{\text{g}}$.\cite{Schmidtke_PhysRevE_2012} This indicates that the energy scale of the high-temperature dynamics $E_{\infty}$ may be relevant for the glassy slowdown and the question arises whether there are theoretical hints for a connection between high- and low-temperature dynamics. This implies the necessity of a model describing the glassy slowdown of glass-forming liquids over a wide temperature range. 

While there are numerous theories describing the dynamics of glass-forming liquids,\cite{Adam_JCP_1965,Kirkpatrick_PhysRevA_1989,Royall_PhysRep_2015,Dyre_JNCS_2006,Dyre_RevModPhys_2006,Simmons_SoftMatter_2012} there is no generally accepted theory to predict the temperature dependence of diffusion coefficients and relaxation times. For example, the mode-coupling theory (MCT) works well in the weakly supercooled regime but increasingly deviates from experiments near $T_{\text{g}}$. Also empirical models like the Vogel-Fulcher-Tammann (VFT) equation or the ansatz from Mauro et al. \cite{Mauro_PNAS_2009} fit to experiments only within a limited temperature range.\cite{Course5_Kob_2003,Schmidtke_JCP_2013} In the simple-liquid regime at high temperatures, in turn, the temperature dependence of the viscosity $\eta$ can be described by the Arrhenius equation $\eta(T)=\eta_{\infty}\exp\left\lbrace E_{\infty}/(k_{\text{B}}T)\right\rbrace$ with the temperature-independent activation energy $E_{\infty}$ \cite{Hrma_JNonChrystSolids_2008,Messaadi_JChem_2015} (in the following, we set $k_{\text{B}}=1$).

Recently, Schmidtke et al. \cite{Schmidtke_PhysRevE_2012} introduced an empirical ansatz for the rotational relaxation time $\tau_{\text{rot}}$ with a temperature-dependent activation energy which is composed of a temperature-independent part equal to $E_{\infty}$ from the high-temperature Arrhenius equation and a temperature-dependent part related to cooperative dynamics at low temperatures. Altogether, it reads $\tau_{\text{rot}}(T)=\tau_{\infty}\exp\left\lbrace\left(E_{\infty}+E_{\text{coop}}(T)\right)/T\right\rbrace$. This ansatz may be related to the ECNLE (elastically collective non-linear Langevin equation) approach \cite{Mirigian_PhysChemLett_2013,Mirigian_JCP_2014_I,Mirigian_JCP_2014_II} that combines MCT calculations and elastic models in a nonlinear Langevin equation framework. The ECNLE approach provides a two-barrier picture of the glassy slowdown with a local and a collective barrier whereat the latter becomes dominant close to $T_{\text{g}}$. The cooperative energy $E_{\text{coop}}(T)$ can be obtained from $\tau(T)$ based on $E_{\text{coop}}(T)=T\ln\left(\tau/\tau_{\infty}\right)-E_{\infty}$ by determining $E_{\infty}$ and $\tau_{\infty}$ from an Arrhenius fit to the high-temperature regime. Moreover, Schmidtke et al. proposed that the cooperative contribution to the activation energy has an exponential temperature dependence
\begin{equation}
	E_{\text{coop}}(T)=E_{\infty}\exp\left\lbrace-\mu\left(\frac{T-T_{\text{A}}}{E_\infty}\right)\right\rbrace
	\label{eq:SR_Ecoop}
\end{equation}
with the generalized fragility $\mu$, high-temperature activation energy $E_{\infty}$, and onset temperature $T_{\text{A}}$ at which the cooperative energy becomes equal to $E_{\infty}$.\cite{Schmidtke_PhysRevE_2012} Such exponential temperature dependence was found based on rotational relaxation times from experimental studies of a large number of molecular liquids \cite{Schmidtke_JCP_2013,Schmidtke_PhysRevE_2012} and polymer melts,\cite{Schmidtke_MM_2015} and from molecular dynamics (MD) simulations of water \cite{Horstmann_JCP_2017} and ionic liquids.\cite{Pal_JCP_2019} Moreover, previous works have suggested \text{$T_{\text{A}}\approx 0.104\times E_{\infty}$} for molecular liquids.\cite{Schmidtke_PhysRevE_2012,Schmidtke_JCP_2013,Schmidtke_MM_2015} Schmidtke's empirical model provides a good description of rotational relaxation times over a large temperature range from the high-temperature regime down to the glass transition. When $E_{\infty}$ and $\tau_{\infty}$ are determined from liquid theories (e.g., the Enskog theory \cite{Chapman_Book_MathematicalTheoryOfNonUniformGases_1970}) and the phenomenological relation \text{$T_{\text{A}}\approx 0.104\times E_{\infty}$} is employed, there is only one free parameter $\mu$ left to describe the glassy slowdown. Additionally, experimental and computational approaches to the rotational motion of supercooled liquids reported common ratios \text{$E_{\infty}/T_{\text{g}}$}. While \text{$E_{\infty}/T_{\text{g}}\approx11$} was reported for a large number of molecular liquids,\cite{Schmidtke_PhysRevE_2012,Horstmann_JCP_2017,Pal_JCP_2019} a ratio of \text{$E_{\infty}/T_{\text{g}}\approx16$} was found for polymers.\cite{Schmidtke_MM_2015} Therefore, it is an intriguing question whether a universal ratio \text{$E_{\infty}/T_{\text{g}}$} exists or whether at least subclasses of glass-forming liquids show common ratios. If so, the high-temperature quantity $E_{\infty}$ would essentially allow one to predict the glass transition temperature.

In this work, we modify the empirical ansatz of Schmidtke et al. in two important aspects. First, we consider translational rather than rotational motion. Second, we assume that the approach applies to the shear viscosity rather than the relaxation time. Specifically, the Stokes-Einstein (SE) relation \text{$D\eta=k_{\text{B}}T/(6\pi R)$} \cite{Book_TheoryOfSimpleLiquids_Hansen_2006} predicts the relation \text{$D\propto T/\eta$} for the translational self-diffusion coefficient $D$. If we assume an Arrhenius-like temperature dependence of the viscosity $\eta$ at high temperatures,\cite{Hrma_JNonChrystSolids_2008,Messaadi_JChem_2015} the temperature dependence of $D$ will be given by $1/D=1/(\xi_DT)\exp\left\lbrace E_{\infty}/T\right\rbrace$ with a constant $\xi_D$. The structural relaxation time $\tau$ is related to $D$ by \text{$\tau\propto 1/D$} in the simple-liquid regime. This can be determined from the self part of the intermediate scattering function in the hydrodynamic limit $F_{\text{s}}(k,t)\propto\exp\left\lbrace -k^2Dt\right\rbrace\propto\exp\left\lbrace-t/\tau\right\rbrace$.\cite{Book_TheoryOfSimpleLiquids_Hansen_2006,Shi_JCP_2013} In order to be consistent with the SE relation and an Arrhenius temperature dependence of $\eta$ at high temperatures, our modified empirical model is given by
\begin{equation}
	\tau(T)=\frac{\xi_{\tau}}{T}\exp\left\lbrace\frac{E_{\infty}+E_{\text{coop}}(T)}{T}\right\rbrace.
	\label{eq:CTB_tau}
\end{equation}
Note that the factor $1/T$ is relevant at high temperatures, in particular for the value of $E_{\infty}$, but only slightly affects the low-temperature behavior. In the following, the ansatz given by Eq.~(\ref{eq:CTB_tau}) will be referred to as cooperative two-barrier (CTB) model.

We use MD simulations to test the applicability of the CTB model. While previous MD simulation works, which have investigated glassy dynamics, explored water-like and silica-like models,\cite{Geske_ZPhysChem_2018,Klameth_JCP_2013,Klameth_PhysChemLett_2015,Horstmann_JCP_2017,Horstmann_JCP_2019,Geske_JCP_2017,Geske_AIPAdv_2016,Sanjon_JCP_2018} which form tetrahedral networks and show various anomalies,\cite{Horstmann_JCP_2017,Horstmann_JCP_2019} we focus on simpler glass-forming mixtures in order to get further insights into the basic mechanisms of the glass transition in the framework of the CTB model. Specifically, we consider binary mixtures which are based on the time-honored Kob-Andersen (KA) system.\cite{Kob_PhysRevLett_1994,Kob_PhysRevE_1995} The KA system consists of two particle species (A and B with $80\,\%$ A and $20\,\%$ B particles) interacting with the Lennard-Jones (LJ) potential. It was studied in numerous works \cite{Geske_ZPhysChem_2018,Kob_PhysRevE_1995,Kob_PhysRevLett_1994,Pedersen_PhysRevLett_2010,Pedersen_PhysRevLett_2018,Ingebrigtsen_PhysRevX_2019,Flenner_PhysRevE_2005,Bruening_PhysCondMatt_2009,Fernandez_PhysRevE_2003,Toxvaerd_JCP_2009,Weber_PhysRevB_1985,Wahnstrom_PhysRevA_1991,Glotzer_JNCS_2000,Heuer_PhysRevE_2005,Nave_PhysRevE_2006,Kob_PhysRevLett_1997,Donati_PhysRevE_1999,Donati_PhysRevLett_1998,Vogel_JCP_2004} reporting a critical temperature \text{$T_{\text{C}}^*=0.435$} from MCT analyses \cite{Kob_PhysRevLett_1994,Kob_PhysRevE_1995} as well as heterogeneous and cooperative dynamics.\cite{Glotzer_JNCS_2000,Heuer_PhysRevE_2005,Nave_PhysRevE_2006,Kob_PhysRevLett_1997,Donati_PhysRevE_1999,Donati_PhysRevLett_1998,Vogel_JCP_2004}

Here, we investigate whether the CTB model can describe the glassy slowdown of KA systems. To take a broad perspective, we add partial charges to the particles. For such charged particles, the interactions and dynamics of the system can be strongly and systematically altered by varying the values of the assigned partial charges, which in turn allows one to scrutinize proposed universalities, as was demonstrated in previous studies of water-like and silica-like systems.\cite{Horstmann_JCP_2017,Sanjon_JCP_2018} The intention of the present work is to test the existence of a generic relation between $E_{\infty}$ and $T_{\text{g}}$, and to test the universality of the CTB model. Explicitly, considering that the KA system was parametrized to model metallic glass formers, one may expect that our study yields insights into the applicability of the CTB model to this important class of materials. 

As our main results, we show that both $E_{\infty}$ and $T_{\text{g}}$ scale quadratically with the charges of the particles independently of the particle species. Explicitly, the ratio \text{$E_{\infty}/T_{\text{g}}$} is approximately constant for all charge values indicating the existence of a generic link between high-temperature dynamics and the glass transition. Remarkably, we demonstrate that the CTB model describes dynamical quantities of our simple glass-forming mixtures over the whole temperature range.

\section{Simulation Details}
\label{sec:SimulationDetails}

\begin{figure*}
	\centering
	\includegraphics[width=0.8\linewidth]{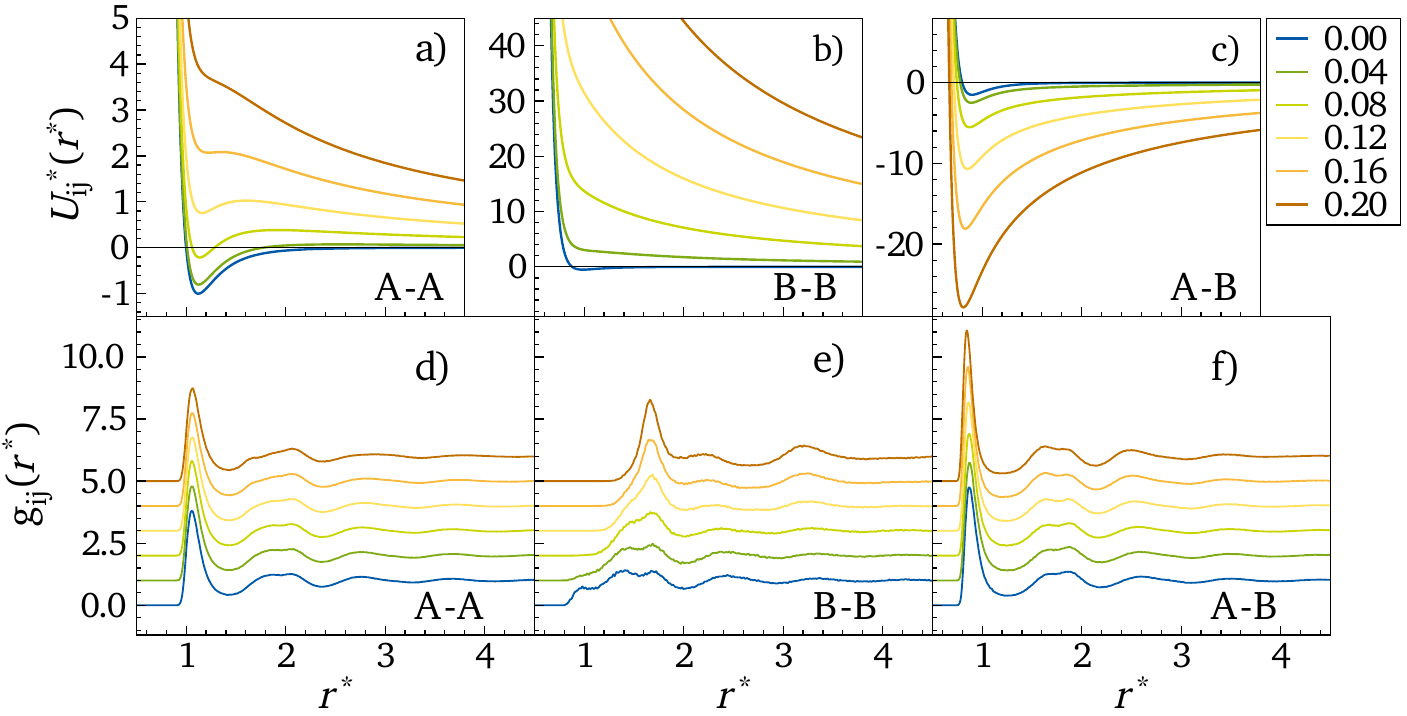}
	\caption[Pair potentials and radial pair-distribution functions]{a)--c) Pair potentials of the charged KA systems. The respective values of the partial charges $q_{\text{A}}^*$ are indicated. d)--f) Corresponding partial RDFs $g_{\text{AA}}$, $g_{\text{BB}}$ and $g_{\text{AB}}$ obtained in the NVT ensemble with \text{$\rho^*=1.2$} at \text{$T^*=0.65$}. The curves are shifted vertically by \text{$n=0,1,2,\dots$} in ascending order with \text{$n=0$} for \text{$q_{\text{A}}^*=0.00$}, and the data is smoothed with a low-pass filter. The corresponding data for the charged KA systems in the NpT ensemble can be found in Fig.\ S1 of the supplementary material. The temperature-dependent RDFs for the uncharged KA system ($q_{\text{A}}^*=0.00$) in both ensembles are compiled in Fig.\ S2.}
	\label{fig:fig1potrdf}
\end{figure*}

The investigated systems are based on the KA liquid which consists of $80\,\%$ A and $20\,\%$ B particles. These have the same mass \text{$m=1\,\text{u}$} and the interaction is modeled by the LJ potential with effective particle diameters \text{$\sigma_{\text{AB}}/\sigma_{\text{AA}}=0.8$}, \text{$\sigma_{\text{BB}}/\sigma_{\text{AA}}=0.88$} and depths of the potential minima characterized by \text{$\epsilon_{\text{AB}}/\epsilon_{\text{AA}}=1.5$} and \text{$\epsilon_{\text{BB}}/\epsilon_{\text{AA}}=0.5$}. All units are scaled with LJ units resulting in a reduced unit system with length \text{$r^*=r/\sigma_{\text{AA}}$}, time \text{$t^*=t/\sqrt{m\sigma_{\text{AA}}^2/\epsilon_{\text{AA}}}$}, charge \text{$q^*=q/e$}, temperature \text{$T^*=k_{\text{B}}T/\epsilon_{\text{AA}}$}, energy \text{$E^*=E/\epsilon_{\text{AA}}$}, and pressure \text{$p^*=p\sigma_{\text{AA}}^3/\epsilon_{\text{AA}}$}. Adding partial charges to the particles, the overall interaction potential between particle $i$ and $j$ is given by
\begin{equation}
	U_{ij}\left(r_{ij}\right)=4\epsilon_{ij}\left[\left(\frac{\sigma_{ij}}{r_{ij}}\right)^{12} - \left(\frac{\sigma_{ij}}{r_{ij}}\right)^6 \right]+\frac{q_iq_j}{4\pi\epsilon_0 r_{ij}}
	\label{eq:InteractionPotential}
\end{equation}
with \text{$r_{ij}=\left|\vec{r}_i-\vec{r}_j\right|$}. The reduced charge $q_{\text{A}}^*$ of the A particles is varied from 0.00 to 0.20 in steps of 0.02. For the B particles, the charge is given by \text{$q_{\text{B}}^*=-4 q_{\text{A}}^*$} such that the entire system is charge neutral. In Fig.~\ref{fig:fig1potrdf} a)--c), we see that the minima in $U_{\text{AA}}$ become more shallow when $q_{\text{A}}^*$ is increased and disappear near $q_{\text{A}}^*=0.2$, while the minimum of $U_{\text{BB}}$ vanishes already near $q_{\text{A}}^*=0.04$ owing to the stronger electrostatic repulsion between the B particles. The strong dependence of the interaction potentials on the value of $q_{\text{A}}$, on the one hand, indicates that our charge scaling approach allows us to significantly alter the dynamics of the studied binary mixtures, and on the other hand, prevents us from using even higher partial charges. 

For the calculation of the pair interactions, the short-ranged LJ potential is truncated and shifted to zero at a cutoff radius of \text{$r_{\text{C}}^*=3.0$}. For the calculation of the long-ranged Coulomb interactions, the particle-mesh Ewald (PME) method is used.\cite{Darden_JCP_1993} Thus, the Coulomb interactions are calculated exactly within the cutoff radius $r_{\text{C}}^*$, and beyond PME is used with a Fourier spacing of \text{$r_{\text{F}}^*=0.3$}. 

All systems consist of 3200 A particles and 800 B particles. They are simulated with a time step of \text{$\Delta t^*=0.004$} under isochoric conditions with constant number density \text{$\rho^*=1.2$} (NVT ensemble) and under isobaric conditions with constant pressure \text{$p^*=0.0602$} (NpT ensemble). While the former ensemble was used in most previous MD simulations of the KA system, the latter is closer to the experimental situation and was also used in the computational analysis of molecular systems.\cite{Horstmann_JCP_2017,Pal_JCP_2019} We use the GROMACS MD simulation package \cite{Abraham_SoftwareX_2015} and perform classical MD simulations in a cubic box with periodic boundary conditions. For the temperature coupling, we employ the velocity-rescaling thermostat.\cite{Bussi_JCP_2007} The pressure coupling in the NpT ensemble is done with the Parrinello-Rahman barostat \cite{Parrinello_ApplPhys_1981} and the Berendsen barostat \cite{Berendsen_JCP_1984} for equilibration purposes. Prior to the production runs, we performed sufficiently long equilibration runs which also serve to adjust the volume and the density in the simulations in the NpT ensemble. All systems are assumed to be equilibrated when the average particle displacements exceed the next-neighbor distances and the intermediate scattering function has decayed to zero, which is achieved by using equilibration times of at least $100\tau_{\text{ISF}}$. We investigate all systems at 30 temperature values equally spaced on the \text{$1/T^*$}-scale from \text{$T^*=0.4$} to \text{$T^*=5.0$} in the NVT ensemble and from \text{$T^*=0.36$} to \text{$T^*=1.09$} in the NpT ensemble where studies at higher temperatures are hampered by an onset of evaporation.

\section{Results}
\label{sec:Results}

In the following, we mainly present the results of our simulations in the NVT ensemble while those in the NpT ensemble are qualitatively similar and are given in the supplementary material.

\subsection{Structural Properties}
\label{sub:Structure}

For a first characterization of the charge-scaled binary mixtures, we investigate their local structure. For this purpose, we analyze the radial pair-distribution functions (RDF)
\begin{equation}
	g_{ij}(r) = \frac{1}{\rho_j N_i}\sum\limits_{k\in S_i}\sum\limits_{\substack{l\in S_j \\ l\neq k}}\left\langle\frac{\delta\left(r-\left|\vec{r}_k(t)-\vec{r}_l(t)\right|\right)}{4\pi r^2}\right\rangle_t
	\label{eq:RDF_calc}
\end{equation}
with $S_i$ and $S_j$ denoting the sets of all particles of type $i$ and $j$, respectively. Moreover, $N_i$ is the total number of particles of species $i$, $\rho_j$ the average number density of particle species $j$, and $\left<\cdot\right>_t$ denotes a time average.

The effect of the charge scaling on the partial RDFs at a fixed temperature is demonstrated in Fig.~\ref{fig:fig1potrdf} d)--f) showing $g_{\text{AA}}$, $g_{\text{BB}}$ and $g_{\text{AB}}$, respectively. Whereas $g_{\text{AA}}$ and $g_{\text{AB}}$ weakly depend on the value of the partial charge, the charge scaling has a strong effect on $g_{\text{BB}}$. For \text{$q_{\text{A}}^*=0.00$}, the B particles have a complex local environment. A shoulder at \text{$r^*\approx0.9$} indicates the next-neighbor distance, and the existence of two further peaks at \text{$r^*\approx1.4$} and 1.7 indicates the absence of well defined coordination shells. When the partial charge is increased, these features merge into a single next-neighbor peak. For \text{$q_{\text{A}}^*=0.20$}, the next-neighbor peak occurs at a large distance of \text{$r^*\approx1.7$}. These effects can be rationalized based on the fact that the electrostatic repulsion is much stronger between the B particles than between the A particles because of \text{$|q_{\text{B}}^*|=4\times|q_{\text{A}}^*|$}.

The important conclusion from Fig.~\ref{fig:fig1potrdf} is that all systems are amorphous independently of the charge values. Therefore, we do not find evidence that additional Coulomb interactions facilitate crystallization of the binary KA mixture. Precisely, the electrostatic interactions suppress a demixing of the system due to the strong repulsion of the B particles which interferes with the formation of a face-centered cubic crystal of A particles, which is crucial to the crystallization of the system,\cite{Fernandez_PhysRevE_2003,Ingebrigtsen_PhysRevX_2019} and hence, improve its glass-forming ability.

\subsection{Dynamical Properties}
\label{sub:Dynamics} 

In order to investigate the glassy slowdown, we analyze the dynamics both at diffusive and molecular length scales. These are probed by the mean square displacement (MSD) and the self part of the intermediate scattering function (ISF), respectively. 

\begin{figure}
	\centering
	\includegraphics[width=0.8\linewidth]{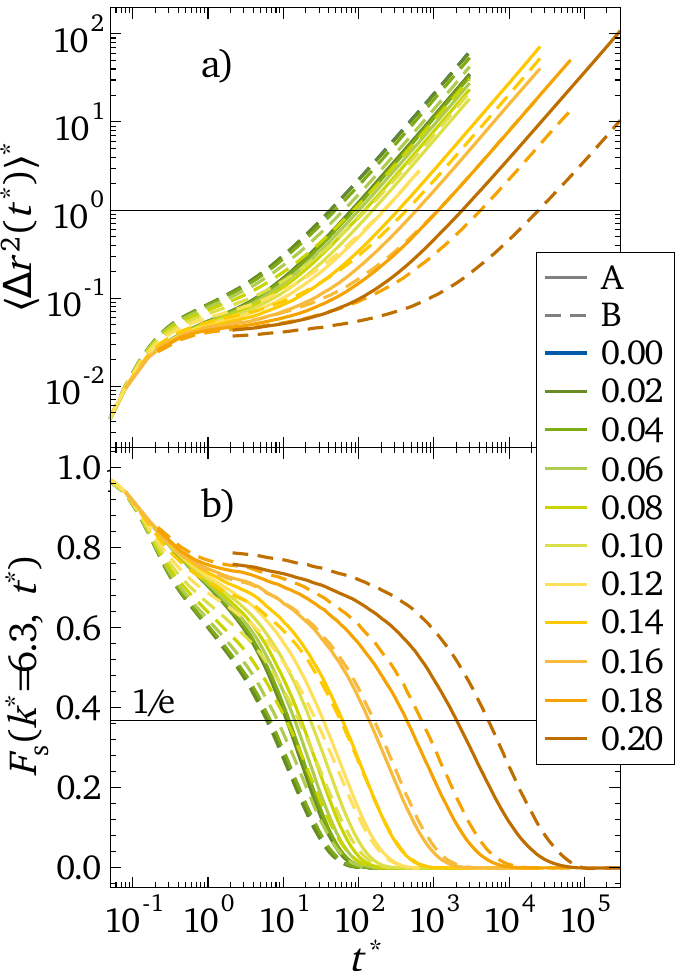}
	\caption[MSD and ISF]{a) Mean square displacement (MSD) of the A (solid lines) and B (dashed lines) particles in the charged KA systems in the NVT ensemble at a fixed temperature \text{$T^*=0.65$}. The colors distinguish between the charges $q_{\text{A}}^*$ with values given in the legend. The black horizontal solid line indicates the MSD corresponding to the squared diameter of the A particles. b) Intermediate scattering function (ISF) of the same systems at \text{$T^*=0.65$} with \text{$k^*=6.3=2\pi/\sigma_{\text{AA}}^*$}. The black horizontal solid line indicates the value $1/e$ at which $\tau_{\text{ISF}}^*$ is determined.}
	\label{fig:fig2msdisf}
\end{figure}

The MSD is a measure for the average quadratic distance a particle moves away from the initial position $\vec{r}\left(t_0\right)$ during the time $t$. In our binary systems, it is given by
\begin{equation}
	\left<\varDelta\vec{r}_p^{~2}(t)\right>=\frac{1}{N_p}\left\langle\sum_{i\in S_p}\left(\vec{r}_i\left(t_0+t\right)-\vec{r}_i\left(t_0\right)\right)^2\right\rangle_{t_0},
	\label{eq:MSDCalc} 
\end{equation}
where $p$ denotes the particle species, $S_p$ represents particles belonging to species $p$, and $N_p$ is the total number of $p$ particles. In Fig.~\ref{fig:fig2msdisf} a), the MSD is exemplarily shown for the A and B particles in the NVT ensemble at a fixed temperature \text{$T^*=0.65$}. At short times, a ballistic regime can be observed whereas a plateau arises at intermediate times due to the cage effect.\cite{Course5_Kob_2003} The long-time behavior of the MSD, i.e., the diffusive regime, is related to the self-diffusion coefficient $D$ by 
\begin{equation}
	\left<\varDelta\vec{r}^{~2}(t)\right>=6Dt
	\label{eq:MSD_D}
\end{equation}
in the case of a three-dimensional system. The black horizontal solid line in Fig.~\ref{fig:fig2msdisf} a) indicates the value \text{$\left<\varDelta\vec{r}^{~2}\right>^*=1.0$} corresponding to the squared diameter of the A particles. The diffusion coefficient $D^*$ is obtained from the MSD by a fit of Eq.~(\ref{eq:MSD_D}) to the diffusive regime at large $t^*$. For both A and B particles, the MSD curves strongly shift to larger $t^*$ for increasing partial charges, and hence, the considered electrostatic interactions slow down the diffusive dynamics of both particle species by several orders of magnitude.

For the investigation of the dynamics at molecular length scales, we use the ISF. Since we have isotropic systems, its self part is given by
\begin{equation}
	F_{\text{s},p}(k,t)=\frac{1}{N_p}\left\langle\sum_{j\in S_p}\frac{\sin\left(k\left|\vec{r}_j\left(t_0+t\right)-\vec{r}_j\left(t_0\right)\right|\right)}{k\left|\vec{r}_j\left(t_0+t\right)-\vec{r}_j\left(t_0\right)\right|}\right\rangle_{t_0}
	\label{eq:Isotropic_ISF}
\end{equation}
and measures how strongly the molecular configuration of the system at time $t_0$ differs from the configuration at time \text{$t_0+t$} at the inverse length scale $k$. In this work, we use \text{$k^*=2\pi/\sigma_{\text{AA}}^*=6.3$} corresponding to the diameter \text{$\sigma_{\text{AA}}^*=1.0$} of the A particles. Hence, the ISF measures the structural relaxation at typical next-neighbor distances. In Fig.~\ref{fig:fig2msdisf} b), it is exemplarily shown for the A and B particles in the NVT ensemble at a fixed temperature \text{$T^*=0.65$}. Similar to the MSD, a plateau arises at intermediate times due to the cage effect before the ISF decays to zero in the structural $\alpha$ relaxation. To extract the $\alpha$-relaxation time $\tau_{\text{ISF}}^*$, we use the criterion \text{$F_{\text{s}}\left(k^*=6.3,t^*=\tau_{\text{ISF}}^*\right)=1/e$}. Similar to the diffusive motion, the $\alpha$ relaxation slows down by more than two orders of magnitude when $q_{\text{A}}^*$ is increased from 0.00 to 0.20. This general slowdown is caused by the charge-dependent attractive interaction between A and B particles.

In addition to a strong slowdown of dynamics, a dynamical decoupling between A and B particles can be observed with increasing $q_{\text{A}}^*$. For \text{$q_{\text{A}}^*<0.10$}, the MSD of the B particles is larger than that of the A particles (cf. Fig.~\ref{fig:fig2msdisf} a). For larger $q_{\text{A}}^*$, this relation is reversed. Qualitatively the same effects can be observed at any other temperature. Thus, the B particles diffuse faster than the A particles in the weakly-charged cases and vice versa in the strongly-charged cases. A similar effect can be observed in the ISF (cf. Fig.~\ref{fig:fig2msdisf} b). For \text{$q_{\text{A}}^*<0.10$}, the ISF decays faster to zero for the B particles than for the A particles and vice versa for \text{$q_{\text{A}}^*>0.10$}. For \text{$q_{\text{A}}^*=0.10$}, the dynamics of A and B particles take place at approximately the same time scales. Thus, the scaling of charges allows us to adjust the relative mobility of the components of the binary mixture. The dynamical decoupling can be rationalized based on the interaction potentials of the particles (cf. Fig.~\ref{fig:fig1potrdf} a--c): In terms of the LJ parameters, the B particles are smaller than the A particles and have weaker interactions among each other, resulting in faster dynamics. However, the additional Coulomb interaction slows down the dynamics of the B particles much stronger than that of the A particles since it is 16 times larger (\text{$q_{\text{B}}=-4q_{\text{A}}$}). Thus, the charge scaling causes non-trivial changes of the dynamics of A and B particles, and therefore, facilitates the analysis of general phenomena related to the glassy slowdown.

The above findings for the dynamics of the binary mixtures imply that $E_{\infty}$ and $T_{\text{g}}$ exhibit a strong charge dependence. Therefore, we analyze the temperature dependence of $D^*$ and $\tau_{\text{ISF}}^*$. In the following, we focus on the results obtained for the MSD in the NVT ensemble. Analogous analyses of the MSD in the NpT ensemble and of the ISF in both ensembles can be found in the supplementary material (Figs.\ S3--S7).

\subsection{Cooperative Two-Barrier Model for the Glassy Slowdown}
\label{sub:CTBAnalysis}

First, we focus on the high-temperature regime. The high-temperature activation energy $E_{\infty}$ can be obtained from a fit of the modified Arrhenius equation $1/D=1/(\xi_DT)\exp\left\lbrace E_{\infty}/T\right\rbrace$ to the high-temperature regime of the self-diffusion coefficient $D$. The fit describes our simulation data in the high-temperature regime almost perfectly. This is shown for the uncharged mixture in Fig.~\ref{fig:fig3tauecoop} a) and for the charged ones in Fig.\ S3 in the supplementary material. Moreover, we observe that $\tau_{\text{ISF}}(T)=\xi_{\tau}/T\exp\left\lbrace E_{\infty}/T\right\rbrace$ is valid for all studied mixtures at sufficiently high temperatures. The finding that $1/D$ and $\tau_{\text{ISF}}$ obey the modified Arrhenius equations is particularly evident in the NVT ensemble where evaporation does not interfere with studies at high temperatures. It indicates that $\eta$ rather than $D$ or $\tau$ obey the Arrhenius law in the simple-liquid regime, and thus, confirms the necessity of our modification of the previous approach by Schmidtke et al.\cite{Schmidtke_PhysRevE_2012} Similarly, a fit of the CTB model $1/D=1/(\xi_DT)\exp\left\lbrace\left(E_{\infty}+E_{\text{coop}}(T)\right)/T\right\rbrace$, which will be discussed in more detail below, gives values for $E_{\infty}$ and $\xi_D$. As we show in Fig.\ S4 a) and b) in the supplementary material, both fit approaches yield consistent results. In the following, we use the results of the fits with the CTB model for all further analysis. We find that, for both particle species, $E_{\infty}$ exhibits a quadratic charge dependence of the form \text{$E_{\infty}\left(q_{\text{A}}\right)=aq_{\text{A}}^2+b$}, see Fig.~\ref{fig:fig4EinfTgChargeDependence} a). Thus, $E_{\infty}$ has a constant contribution $b$ from the LJ potential and a charge-dependent part that scales like the Coulomb interaction. The values of the corresponding dimensionless fit parameters are given by \text{$a^*=9.00\pm0.86$} and \text{$b^*=1.58\pm0.02$} for the A particles, and \text{$a^*=28.6\pm0.11$} and \text{$b^*=1.48\pm0.02$} for the B particles. As expected from the larger charges of the B particles, their activation energy has a stronger charge dependence. Explicitly, when $q_{\text{A}}^*$ is increased from 0.00 to 0.20, $E_{\infty}^*$ grows by a factor of about 1.2 from 1.6 to 1.9 for the A particles and by a factor of about 1.9 from 1.4 to 2.6 for the B particles. Without our modification of the Arrhenius equation, one would get activation energies that are higher by up to a factor of two, see Fig.\ S5 a) in the supplementary material.

Next, we analyze whether the CTB model allows us to describe our data over the whole temperature range. We see in Fig.~\ref{fig:fig3tauecoop} a) for the diffusion coefficients of the uncharged A particles in the NVT ensemble that the CTB model describes not only the high-temperature behavior, but also the non-Arrhenius temperature dependence in the supercooled regime. Moreover, we observe in Fig.~\ref{fig:fig3tauecoop} c) that the CTB model also works for the diffusion of the A particles in the charged KA systems simulated in the NVT ensemble. Additionally, we show in the supplementary material that this approach interpolates the diffusion data of the B particles in the NVT ensemble and of both A and B particles in the NpT ensemble (cf. Fig.\ S6). Finally, the CTB model also describes the temperature-dependent correlation times $\tau_{\text{ISF}}(T)$ for both components and ensembles (cf. Fig.\ S7 in the supplementary material). 

To further scrutinize the validity of the CTB model, we consider the cooperative energy calculated from $E_{\text{coop}}(T)=T\ln\left(T\xi_D/D\right)-E_{\infty}$. It is shown in Fig.~\ref{fig:fig3tauecoop} d) in a generalized representation according to Eq.~(\ref{eq:SR_Ecoop}). We see that the diffusion data of all systems collapse onto a master curve. Here, $E_{\text{coop}}(T)$ shows the exponential temperature dependence (black lines) proposed by the CTB model. Merely at very small values of $E_{\text{coop}}$, statistical uncertainties of the data cause some scattering of the values. Thus, the slowdown of the diffusion can be traced back to an exponential temperature dependence of the cooperative energy for all studied systems. This result also applies to the structural relaxation time as shown in Fig.~\ref{fig:fig3tauecoop} d).

The generalized fragility $\mu$ varies between 10 and 22 for the A and B particles in the NVT ensemble. However, we did not find any correlation with the partial charges. In contrast, $T_{\text{A}}$ exhibits a quadratic charge dependence. A parabolic fit of the form \text{$T_{\text{A}}(q_{\text{A}})=aq_{\text{A}}^2+b$} yields the reduced parameters \text{$a^*=5.7\pm0.2$} and \text{$b^*=0.452\pm0.004$} for the A particles, and \text{$a^*=6.1\pm0.2$} and \text{$b^*=0.440\pm0.004$} for the B particles in the NVT ensemble. Increasing $q_{\text{A}}^*$ from 0.00 to 0.20 increases $T_{\text{A}}^*$ from 0.43 to 0.67. The ratio \text{$T_{\text{A}}^*/E_{\infty}^*$} is about 0.3 for the A and B particles in the NVT ensemble and does not depend on $q_{\text{A}}^*$. In the NpT ensemble, the ratio is also independent of the partial charges and amounts to about 0.16, which is close to the value of 0.104 found in experimental studies of molecular glass formers.\cite{Schmidtke_PhysRevE_2012}

\begin{figure*}
	\centering
	\includegraphics[width=0.8\linewidth]{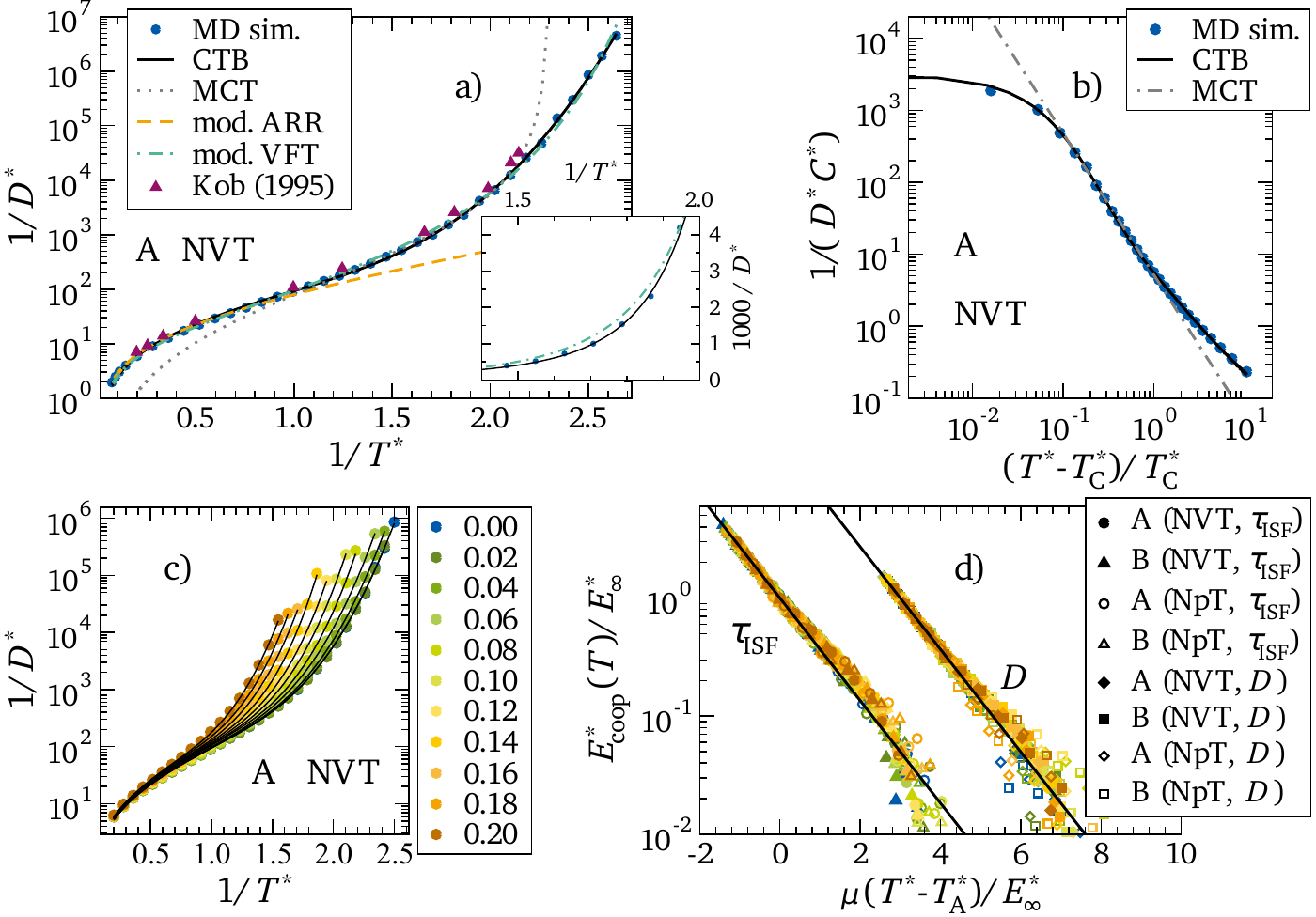}
	\caption[Relaxation times and cooperative energy]{a) Diffusion coefficients of the A particles ($q_{\text{A}}^*=0.00$) in the NVT ensemble with modified Arrhenius, CTB, MCT and modified VFT fits. The purple symbols are data from Ref.\ \onlinecite{Kob_PhysRevE_1995}. The inset shows results at intermediate temperatures demonstrating the difference between the modified VFT fit and the CTB fit. b) The same diffusion coefficients in a generalized MCT representation indicating that the CTB result overlaps with the MCT fit in the linear MCT regime. c) Diffusion coefficients of the A particles in the NVT ensemble for the indicated values of $q_{\text{A}}^*$. The black solid lines are CTB fits. d) Cooperative energy for all systems in a generalized representation. The colors are the same as in panel c). The data obtained from the diffusion coefficients are shifted by +3 along the $x$-axis for clarity. The black solid lines indicate the exponential temperature dependence of the CTB model.}
	\label{fig:fig3tauecoop}
\end{figure*}

To further investigate the temperature dependence of the diffusion coefficients in the framework of the CTB model, we compare our results to the commonly used MCT. We fit the MCT power law \text{$1/D=C/(T-T_{\text{C}})^{\gamma}$} to $D$ and $\tau_{\text{ISF}}$ and determine the mode-coupling temperature $T_{\text{C}}$. For \text{$q_{\text{A}}^*=0.00$}, we can reproduce \text{$\gamma=2.4$} and \text{$T_{\text{C}}^*=0.435$} as obtained in Ref.~\onlinecite{Berthier_PhysRevE_2010} for $\tau_{\text{ISF}}$. The critical temperature $T_{\text{C}}$ exhibits a quadratic charge dependence of the form \text{$T_{\text{C}}\left(q_{\text{A}}\right)=aq_{\text{A}}^2+b$} with reduced parameters \text{$a^*=3.64\pm0.16$} and \text{$b^*=0.437\pm0.003$} for the A particles, and \text{$a^*=4.24\pm0.20$} and \text{$b^*=0.435\pm0.004$} for the B particles in the NVT ensemble. However, the MCT prediction only describes our data in a limited temperature regime as can be seen in Fig.~\ref{fig:fig3tauecoop} b). It shows $1/D^*$ scaled by the reduced parameter $C^*$ of the MCT power law as a function of \text{$\left(T^*-T_{\text{C}}^*\right)/T_{\text{C}}^*$} together with the CTB and MCT fit in a double-logarithmic representation. Whereas the CTB model covers the data over the whole temperature range, the MCT power law only describes the data in a limited temperature range. Nonetheless, the CTB model perfectly overlaps in the linear MCT regime of the data with the MCT power law, and thus, is compatible with the predictions of the MCT. Furthermore, MCT provides the critical temperature $T_{\text{C}}$ that can be compared to the high-temperature activation energy $E_{\infty}$ and the glass transition temperature $T_{\text{g}}$, see below. 

As the determination of $T_{\text{g}}$ from our simulation data requires substantial extrapolation, we compare results for two different models of the temperature dependence, the CTB ansatz and the analogously modified VFT relations, $1/D=1/(\xi_DT)\exp\left\lbrace B/(T-T_0)\right\rbrace$ and $\tau_{\text{ISF}}=\xi_{\tau}/T\exp\left\lbrace B/(T-T_0)\right\rbrace$. We find that also the modified VFT relations describe the $1/D$ and $\tau_{\text{ISF}}$ data in the high-temperature regime. However, they perform inferior to describe the slowdown in the weakly supercooled regime, see, e.g., Fig.~\ref{fig:fig3tauecoop} a), which is the main reason for using the CTB model. In order to obtain $T_{\text{g}}$ values from the correlation times, we extrapolate the CTB and modified VFT fits and use the criterion $\tau_{\text{ISF}}^*(T_{\text{g}})=\tau_{\text{g}}^*=10^{15}$, corresponding to the typical value \text{$\tau_{\text{g}}=1000\,\text{s}$} at the glass transition temperature. For the analogous analysis of the diffusion data, we extrapolate these temperature dependencies to \text{$D_{\text{g}}^*\equiv D^*(T_{\text{g}}^*)$}. Here, $D_{\text{g}}^*$ is the diffusion coefficient corresponding to \text{$\tau_{\text{g}}^*=10^{15}$}. To determine the values of $D_{\text{g}}$ for the studied systems, we need to consider, however, the breakdown of the SE relation in the KA system, which can be described by the power-law relation \text{$D\propto (1/\tau)^{\beta}$} (\text{$\beta \leq 1$}).\cite{Bordat_JPhysCondensMatter_2003} Therefore, $D^*(\tau^*)$ is fitted with a power law in the supercooled regime and the obtained power laws are extrapolated to \text{$\tau_{\text{ISF}}^*\left(T_{\text{g}}\right)=\tau_{\text{g}}^*=10^{15}$} to extract \text{$D_{\text{g}}^*\equiv D^*(T_{\text{g}}^*)$}. In our systems, $\beta$ varies between 0.69 and 0.92.

\begin{figure}
	\centering
	\includegraphics[width=0.8\linewidth]{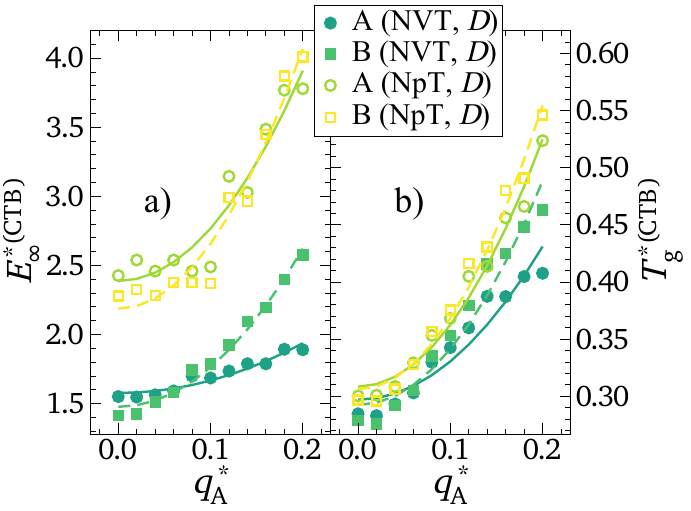}
	\caption{a) Reduced activation energies $E_{\infty}^*$ over reduced charge $q_{\text{A}}^*$ obtained from the CTB analyses of the diffusion data. b) Similar plot for the obtained glass transition temperatures $T_{\text{g}}^*$. Solid lines are parabolic fits.}
	\label{fig:fig4EinfTgChargeDependence}
\end{figure}

The glass transition temperatures obtained from both empirical formulas, the modified VFT equation and the CTB model, are compared in Fig.\ S5 b) in the supplementary material. The ratio $T_{\text{g}}^{*\text{(CTB)}}/T_{\text{g}}^{*\text{(VFT)}}$ is close to one for the diffusion data and lower than one for the structural relaxation time. Hence, the modified VFT equation predicts a stronger slowdown than our CTB model for the structural relaxation. The value of $T_{\text{g}}^{*\text{(CTB)}}/T_{\text{g}}^{*\text{(VFT)}}$ depends on the particle type and the studied observable and varies between 0.8 and 1.2. Therefore, the relative uncertainty in $T_{\text{g}}$ is estimated as approximately $20\,\%$. Remarkably, the ratio does not depend on $q_{\text{A}}^*$, irrespective of the studied observable and the used ensemble. This result indicates that the charge dependence of $T_{\text{g}}$ can be determined in a model-free manner. Both the CTB and the modified VFT fit approaches yield a quadratic charge dependence of $T_{\text{g}}$ that can be described by \text{$T_{\text{g}}\left(q_{\text{A}}\right)=aq_{\text{A}}^2+b$} resembling the findings for $E_{\infty}$, see Fig.~\ref{fig:fig4EinfTgChargeDependence} b). In the CTB analysis of the diffusion data, we find that $T_{\text{g}}^{*\text{(CTB)}}$ rises from about 0.3 to 0.45 and to 0.55 in the NVT and NpT ensembles, respectively, when $q_{\text{A}}^*$ is increased from 0.00 to 0.20, where the rise is somewhat stronger for the B than the A particles. For example, the charge dependencies of $T_{\text{g}}^{*\text{(CTB)}}$ are characterized by the reduced fit parameters \text{$a^*=3.35\pm0.36$} and \text{$b^*=0.30\pm0.01$} for the A particles, and \text{$a^*=4.90\pm0.39$} and \text{$b^*=0.29\pm0.01$} for the B particles. Thus, our charge scaling strongly changes the glass transition temperature.

\begin{figure}
	\centering
	\includegraphics[width=0.8\linewidth]{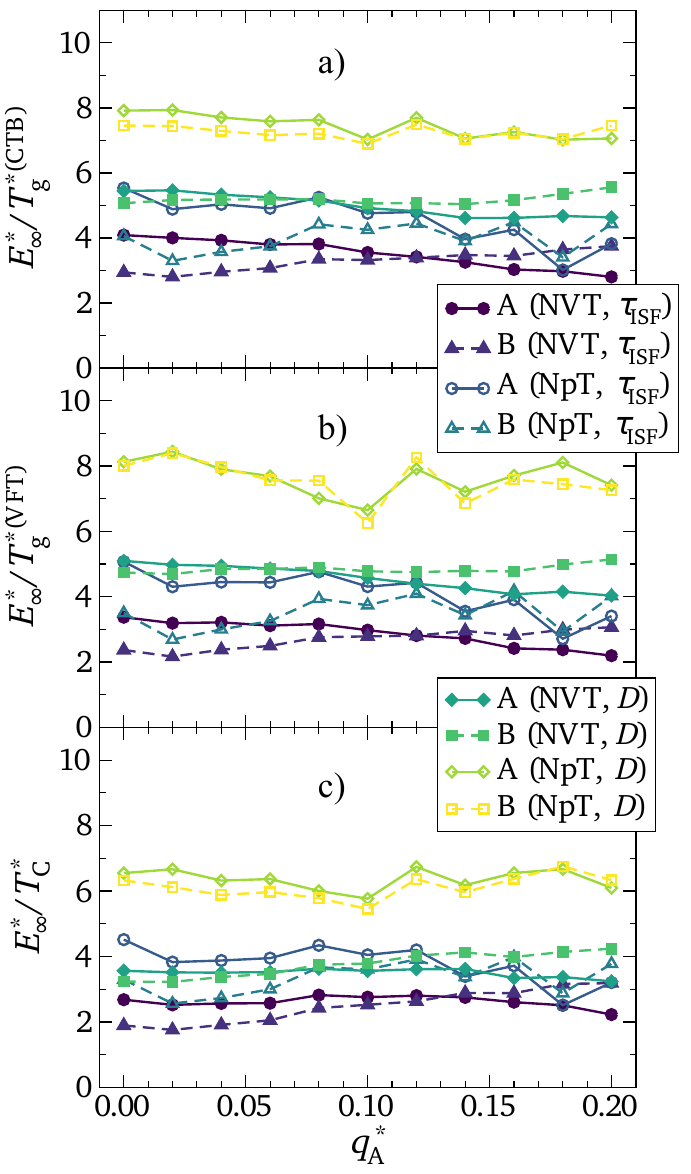}
	\caption[Ratio of high-temperature activation energy and glass transition temperature]{Charge dependence of the ratios \text{$E_{\infty}^*/T_{\text{g}}^{*}$} obtained from a) CTB and b) modified VFT analyses, and c) of ratios \text{$E_{\infty}^*/T_{\text{C}}^{*}$} resulting from MCT fits. Results for the A and B particles, for the NVT and NpT ensembles, and for the diffusion coefficients $D$ and correlation times $\tau_{\text{ISF}}$ are compared in all panels.}
	\label{fig:fig4einftg}
\end{figure}

Finally, we investigate the relation between $E_{\infty}$ and $T_{\text{g}}$. We start with the results from our analysis with the CTB model. Figure~\ref{fig:fig4einftg} a) shows the ratio \text{$E_{\infty}^*/T_{\text{g}}^{*\text{(CTB)}}$} as a function of $q_{\text{A}}^*$. Independently of the ensemble, the observable, and the species, the ratio \text{$E_{\infty}^*/T_{\text{g}}^{*\text{(CTB)}}$} is essentially independent of the charge $q_{\text{A}}^*$ implying a generic relation between high-temperature and low-temperature dynamics for all studied KA mixtures. Moreover, the A and B particles show similar ratios. However, \text{$E_{\infty}^*/T_{\text{g}}^{*\text{(CTB)}}$} is significantly higher for the NpT than for the NVT ensemble. Furthermore, it is larger for $D$ than for $\tau_{\text{ISF}}$ which is related to the different length scales at which $D$ and $\tau_{\text{ISF}}$ probe the dynamics. In particular, we show in Fig.\ S4 c) in the supplementary material that the value of $E_{\infty}^*$ obtained from ISF data for various scattering vectors strongly decreases when $k^*$ is increased in the range \text{$k^*>1$}. Thus, in studies of translational motion, the ratios $E_{\infty}^*/T_{\text{g}}^*$ depend on the length scale of the experiment unless the hydrodynamic limit is reached in the limit of small $k^*$ or, equivalently, diffusion coefficients are analyzed. Qualitatively similar results are obtained when the modified VFT relation is used for the analysis, see Fig.~\ref{fig:fig4einftg} b). However, the ratios $E_{\infty}^*/T_{\text{g}}^{*\text{(VFT)}}$, in general, differ from the ratios $E_{\infty}^*/T_{\text{g}}^{*\text{(CTB)}}$ because of the somewhat different $T_{\text{g}}^*$ values resulting from the modified VFT and CTB extrapolations, see Fig.\ S5 b). The highest ratio $E_{\infty}^*/T_{\text{g}}^*\approx 8$ is obtained for the diffusion data of both A and B particles in the NpT ensemble from the CTB model. However, this value is still smaller than $E_{\infty}/T_{\text{g}}\approx 11$ reported in experimental studies on the rotational motion of molecular liquids.\cite{Schmidtke_PhysRevE_2012}

The existence of a charge-independent relation between $E_{\infty}$ and $T_{\text{g}}$ is confirmed by the results of our MCT analysis. Like $T_{\text{g}}$, the obtained critical temperature $T_{\text{C}}$ exhibits a quadratic charge dependence resulting in a constant ratio \text{$T_{\text{C}}^*/T_{\text{g}}^*$} that is given by \text{$T_{\text{C}}^*/T_{\text{g}}^*\approx 1.4$} in the NVT ensemble. In Fig.~\ref{fig:fig4einftg} c), we see that also the ratio \text{$E_{\infty}^*/T_{\text{C}}^{*}$} is charge independent, further supporting a relation between $E_{\infty}$ and the glassy slowdown. As expected from \text{$T_{\text{C}}>T_{\text{g}}$}, the ratios \text{$E_{\infty}^*/T_{\text{C}}^{*}$} are smaller than the corresponding ratios \text{$E_{\infty}^*/T_{\text{g}}^{*}$} in Fig.~\ref{fig:fig4einftg} a) and b). 

To conclude our analysis on the basis of the CTB model, we apply the model to other common models of glass-forming systems, e.g., the Gaussian core model,\cite{Ikeda_JCP_2011} the harmonic sphere model,\cite{Flenner_JCP_2013} the Wahnström model,\cite{Jenkinson_JCP_2017,Coslovich_PhysRevE_2011} the generalized Hertzian potential (GHP) model,\cite{Miyazaki_JCP_2019} and the hard-sphere model.\cite{Berthier_PhysRevE_2009} We find that also the literature data for these models is well described by the CTB approach, as is shown in Fig.\ S8 of the supplementary material. Moreover, we again determine the ratio between the high-temperature activation energy and the glass transition temperature. We obtain $E_{\infty}^{*}/T_{\text{g}}^*$ ratios which are similar to those of the present charged binary mixtures, see Fig.\ S8 in the supplementary material. Thus, the generic relation between $E_{\infty}$ and $T_{\text{g}}$ applies to a wide variety of glass-forming model liquids.

\subsection{The Role of the Extrapolation}
\label{sub:RoleOfExtrapolation}

\begin{figure}
	\centering
	\includegraphics[width=0.9\linewidth]{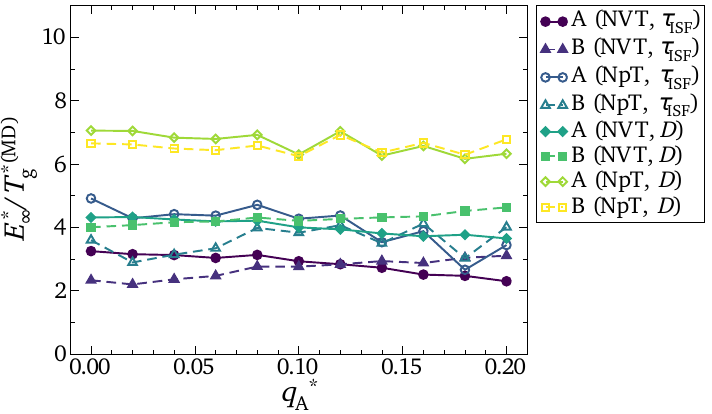}
	\caption{Ratio $E_{\infty}^*/T_{\text{g}}^{*\text{(MD)}}$ with an MD glass transition temperature $T_{\text{g}}^{*\text{(MD)}}$ defined as the temperature at which the structural relaxation time exceeds $\tau^*(T_{\text{g}}^{*\text{(MD)}})=10^7$. The values for $E_{\infty}^*$ and $T_{\text{g}}^{*\text{(MD)}}$ are obtained from fits with the CTB model.}
	\label{fig:fig6}
\end{figure}

To enable straightforward comparison with experimental results, we used the common definition $\tau_{\text{ISF}}(T_{\text{g}})=1000\,\text{s}$ of the glass transition temperature $T_{\text{g}}$ which reads $\tau_{\text{ISF}}^*(T_{\text{g}})=10^{15}$ in reduced units. The determination of $T_{\text{g}}$ then requires extrapolation of the CTB model over approximately ten orders of magnitude in $D^*$ and $\tau_{\text{ISF}}^*$. However, this extrapolation does not influence the qualitative results of our study. To show that, we similarly define an MD glass transition temperature $T_{\text{g}}^{\text{(MD)}}$ as the temperature at which the structural relaxation time exceeds $\tau^*(T_{\text{g}}^{*\text{(MD)}})=10^7$, which means that only a slight extrapolation is necessary to determine $T_{\text{g}}^{\text{(MD)}}$. By using the CTB fits for determining $E_{\infty}^*$ and $T_{\text{g}}^{*\text{(MD)}}$, we obtain qualitatively the same results as with the usual definition of the glass transition temperature, see Fig.\ \ref{fig:fig6}. Again, the ratio $E_{\infty}^*/T_{\text{g}}^{*\text{(MD)}}$ is independent of the charges indicating, that this result does not depend on the actual definition of $T_{\text{g}}$.

\section{Summary and Outlook}
\label{sec:SummaryOutlook}

By performing extensive MD simulations, we have studied various charged glass-forming binary mixtures at temperatures from near the boiling point down to the supercooled regime. The structure of the systems remains amorphous over the whole temperature range but local environments, in particular, the radial pair-distribution function of B particles, change when the partial charges of the particles are varied. Moreover, the charge scaling modifies the time scales of the dynamics by orders of magnitude and alters the relative mobility of the particle species. The temperature dependence of the relaxation times and diffusion coefficients of all investigated systems and both particle species is well described by the CTB model. Thus, the glassy slowdown of all studied mixtures can be traced back to the exponential temperature dependence of the cooperative energy $E_{\text{coop}}(T)$ in the CTB model which should, however, not be mistaken with thermally activated behavior. These findings for the translational motion in the studied charged KA mixtures are consistent with the outcome of previous experimental and computational approaches to the rotational motion of molecular liquids.\cite{Schmidtke_PhysRevE_2012,Schmidtke_JCP_2013,Horstmann_JCP_2017,Schmidtke_MM_2015,Pal_JCP_2019} Furthermore, performing analogous analyses on literature data, we observed that the CTB approach also works for several other common models of glass-forming liquids. Our analyses in extended high-temperature ranges inspire a modification of the approach by Schmidtke et al.,\cite{Schmidtke_PhysRevE_2012,Schmidtke_JCP_2013} who assume that the correlation times $\tau$ (and the diffusion coefficients $D$) obey Arrhenius laws. Specifically, we found that it is more appropriate to assume that the viscosity shows Arrhenius behavior in the high-temperature regime so that, according to the Stokes-Einstein relation, $\tau\times T$ and $T/D$ show this behavior.      

Moreover, we found a common ratio between the high-temperature activation energy $E_{\infty}$ and the glass transition temperature $T_{\text{g}}$ for all charged binary mixtures. However, the value of $E_{\infty}/T_{\text{g}}$ depends on the particle type, the studied observable, and the considered ensemble. In all cases, the ratio is smaller than that observed for the rotational motion in experiments on molecular liquids \cite{Schmidtke_PhysRevE_2012,Schmidtke_JCP_2013} and MD simulations of water \cite{Horstmann_JCP_2017} and ionic liquids.\cite{Pal_JCP_2019} A part of this discrepancy can be related to our modification of the original approach,\cite{Schmidtke_PhysRevE_2012} which results in smaller $E_{\infty}$ values. Moreover, there may be differences between translational and rotational motions. Finally, one should consider that the $T_{\text{g}}$ values of our simulation study result from substantial extrapolation, and hence, are subject to uncertainties, which were quantified by comparing the values from the CTB and modified VFT analyses. Notwithstanding, our result of a constant ratio between $E_{\infty}$ and $T_{\text{g}}$ is robust against the underlying fit model and the actual definition of $T_{\text{g}}$. Moreover, our MCT analyses yielded a constant ratio \text{$E_{\infty}/T_{\text{C}}$} as well.

Our finding of a charge-independent ratio \text{$E_{\infty}/T_{\text{g}}$} is a non-trivial result because the studied mixtures show significant differences in structure and dynamics. Moreover, our CTB analyses for other common model glass formers revealed similar ratios confirming that high-temperature dynamics and the glass transition are linked phenomena. Precisely, using known theories of simple-liquid dynamics and adding one cooperativity parameter, the generalized fragility $\mu$ in the CTB model, allows one to describe the whole glassy slowdown and to predict the glass transition temperature at least for a subclass of glass-forming liquids.

For further analyses, experimental and computational studies should determine the ratio \text{$E_{\infty}/T_{\text{g}}$} for more systems to ascertain which subclasses of glass-forming liquids show common values. A remaining question is whether there is any theory that predicts a constant ratio \text{$E_{\infty}/T_{\text{g}}$} and that provides any physical explanation for an exponential temperature dependence of the cooperative contribution to the activation energy. It is also interesting whether the generalized fragility of the CTB model is connected to any cooperativity or heterogeneity observable like cluster sizes. Remarkably, the electrostatic interactions should improve the glass formation properties of the usual KA system which may be useful for future studies. Due to the strong similarity of our charged binary mixtures with metallic glass-forming liquids, it might also be important to investigate the relation between $E_{\infty}$ and $T_{\text{g}}$ for various metallic glasses. Our results indicate that the prediction of their glass transition temperature might be possible by only determining their activation energy from high-temperature dynamics.

\section*{Supplementary Material}
See the supplementary material for the detailed temperature and charge dependence of the radial pair-distribution functions, the modified high-temperature Arrhenius fits, the CTB fits for all studied systems, ensembles and observables, a comparison of CTB fits with modified Arrhenius fits in the high-temperature regime, a comparison of the glass transition temperatures obtained from the CTB model and the modified VFT relations, the wave number dependence of the high-temperature activation energy obtained from the structural relaxation time, and the ratio between $E_{\infty}$ and $T_{\text{g}}$ for other common models of glass-forming liquids.

\section*{Data Availability}
The data that support the findings of this study are available from the corresponding author upon reasonable request.

\section*{References}
\bibliography{library}

\end{document}